# Comparative study on the uniform energy deposition achievable via optimized plasmonic nanoresonator distributions

M. Csete[1], A. Szenes[1], E. Tóth[1], D. Vass[1], O. Fekete[1], B. Bánhelyi[2],
I. Papp[3,4], T. Bíró[3], L. P. Csernai[3,4,5], N. Kroó[3,6]   (NAPLIFE Collaboration)

[1]*Department of Optics and Quantum Electronics, University of Szeged, 6720 Szeged Hungary*
[2]*Department of Computational Optimization, University of Szeged, 6720 Szeged, Hungary*
[3]*Wigner Research Centre for Physics, Budapest, Hungary*
[4]*Department of Physics and Technology, University of Bergen, 5007 Bergen, Norway*
[5]*Frankfurt Institute for Advanced Studies, 60438 Frankfurt/Main, Germany*
[6]*Hungarian Academy of Sciences, 1051 Budapest, Hungary*

**Abstract**
Plasmonic nanoresonators of core-shell composition and nanorod shape were optimized to tune their absorption cross-section maximum to the central wavelength of a short pulse. Their distribution along a pulse-length scaled target was optimized to maximize the absorptance with the criterion of minimal absorption difference in between neighbouring layers. Successive approximation of layer distributions made it possible to ensure almost uniform deposited energy distribution up until the maximal overlap of two counter-propagating pulses. Based on the larger absorptance and smaller uncertainty in absorptance and energy distribution core-shell nanoresonators override the nanorods. However, optimization of both nanoresonator distributions has potential applications, where efficient and uniform energy deposition is crucial, including phase transitions and even fusion.

**Introduction**
Tailoring the geometry and composition of metal nanoparticles makes it possible to excite localized surface plasmon resonances (LSPRs) at specific frequencies. Two widely studied plasmonic nanoparticle types are the dielectric-metal core-shell nanoparticle and the elongated nanorod. LSPRs excited on core-shell nanoparticles are the hybridized modes of dipolar or higher-order primitive sphere and cavity plasmons [1]. The strength of their interactions depends on the order of both individual modes, the type of hybridization (bonding or anti-bonding) and their interaction-length, i.e. the thickness of the metallic shell. Accordingly, the resonance wavelength of the hybridized modes can be tuned by the generalized aspect ratio (GAR = $r_1/r_2$) of the core-shell particle through a wide spectral range [2, 3]. Another unique property of core-shell nanoparticles is the existence of two distinct GARs, that correspond to nanoresonator compositions exhibiting dipolar resonances at the same frequency in a given medium [4, 5]. The so-called thin shell composition has a larger absorption cross-section due to its small volume, whereas the thick shell composition has significantly larger scattering cross-section.

On metal nanorods the LSPRs can be tuned through a wide spectral range by varying their aspect ratio (AR, which is the ratio of the long-to-short axis) [6, 7]. The advantage of nanorods over core-shell nanoparticles is that their near-field enhancement may be higher due to the large curvature of the metallic apexes, however, their resonant response depends on the relative orientation with respect to the exciting electric field polarization. The longitudinal mode can be excited most efficiently, when the polarization is parallel to the long axis, whereas the transversal resonance appears at smaller wavelengths in the presence of the perpendicular **E**-field component, and is accompanied by smaller absorption cross-section.

The frequency of, the near-field and far-field phenomena accompanying the plasmon resonance in both cases depends on the shape and size, as well as on the material of the nanoresonator and the adjacent medium.

This sensitivity of the resonant frequency to the ambient medium makes plasmonic particles a good candidate in sensing applications [8, 9]. The enhancement of the near-field intensity and directivity of far-field emission in their arrays can be exploited to improve and out-couple the fluorescence of emitters [10, 11]. The relatively small Q-factor accompanied by significant absorption in the metallic segment is advantageous in cancer therapy [12, 13]. The balance of near- and far-field losses is a complex function of the nanoresonator configuration, hence an effective solution to the e.g. maximal energy deposition problem requires numerical optimization.

The absorptance of various targets can be enhanced by embedded plasmonic nanoresonators that can serve as local absorbers throughout their damage threshold. This threshold sensitively depends on the geometry of the nanoresonators. In case of (30 nm diameter) nanospheres a damage threshold of $5*10^{12}$ W/cm$^2$ was reported (800 nm, 30 fs) [14]. In comparison, the highest intensity that might result in LSPR excitation without the damage of nanotriangles is $1.4*10^{12}$ W/cm$^2$ (800 nm, 42 fs) [15]. In case of nanospirals a larger $8*10^{14}$ W/cm$^2$ threshold was described (1500 nm, 20 fs) [15].

Near the damage threshold nonlinear phenomena can arise. The second order nonlinearity ($\chi^{(2)}$) of gold is usually neglected for symmetry reasons, whereas the third order nonlinear susceptibility ($\chi^{(3)}$) strongly depends on the geometry of the nano-object, on the wavelength and pulse-length of the illumination [17]. Full non-perturbative time-domain hydrodynamic description of the electron plasma under femtosecond excitation made possible to describe the nonlinear response of nanostructures. The importance of the resonant effects in the enhancement of the nonlinear interactions as well as the role of topology of the nanostructure was proven [18]. Particles with high spatial symmetry allow nonlinear conversion only between localized plasmon modes with the same symmetry properties. Charge separation accompanying dipoles (quadrupoles) at the second harmonic is smaller (larger) than charge separation arising in presence of the fundamental dipolar mode, however the quadrupolar mode is a dark mode [19].

Coupling nonlinear hydrodynamic equations with the Maxwell equations enabled to describe the boundary second harmonic (SH) response and third harmonic (TH) generation in metallic nanostructures without the need of a priori assumption regarding the nonlinear susceptibility values. From the quadratic and cubic dependence for SH and TH the $\chi^{(2)}$ and $\chi^{(3)}$ were evaluated [20]. Metal is modelled as a two-component medium using the hydrodynamic model to describe free electrons and Lorentz oscillators to account for core electron contributions to both the linear dielectric constant and harmonic generation. It was shown that in case of a silver nanopillars illuminated by 800 nm and 20 fs pulse the transverse magnetic SH (TH) generation is enhanced (decreased) by bound charges [21].

The implantation of plasmonic nanoparticles and nanoshells into fusion ignition targets has been already proposed [22, 23]. The primary studies revealed that the large electromagnetic field enhancement around such nanoresonators is capable of triggering the nuclear chain reactions in conventional deuterated polystyrene fuel targets [22]. It was stated that the coating of the conventional Pd-based fuel materials with nanoparticles and nanoshells has the potential to overcome the Coulomb barriers to produce condensed matter fusion reaction [23]. Later application of ordered nanowire arrays was proposed to transfer matters into the ultra-hot plasma regime with fs laser pulses of moderate energy [24]. Formation of Z-pinches was demonstrated by irradiating nanowires with fs laser pulses of relativistic intensity [25]. The possibility to promote the energy penetration via nanowires and to enhance the plasma extension was also demonstrated [26].

In our previous studies we have proposed a novel approach to promote uniform energy deposition and to ensure time-like ignition of fusion targets by doping them with metal nanoparticles [27]. Energy deposition increased by plasmonic nanoparticles allows the initial compression of the fusion target to be reduced, thus avoiding Rayleigh-Taylor instability that prevents the achievement of Inertial Confinement Fusion [28].

It has also been shown that time-like ignition can be achieved via double sided irradiation with short and intense pulses even at reduced compression, when the absorption of targets is improved by orders of magnitude due to embedded plasmonic core-shell nanoparticles [29]. Beside the energy deposition the charge separation on nanoparticles is also an important aspect, since its amplitude might be commensurate with that accompanying the Laser Wake Field (LWF) dense plasma waves initiated by two intense counter propagating laser pulses [30].

In this study we have exploited the large absorption provided by the dipolar resonance of the thin shell composition of core-shell nanoparticles, as well as the dipolar resonance of nanorods and optimized the geometry of individual nanoresonators as well as their distribution to achieve efficient and uniform energy deposition in fusion targets. The specific purposes were as follows (i) to improve the absorptance of a polymer slab via embedded plasmonic nanoresonators of a core-shell composition and a nanorod shape, and (ii) to determine such distributions that are capable of ensuring uniform energy deposition via short laser pulses of a test-target with a pulse-length scaled size, (iii) to map the charge separation that accompanies resonance near the damage threshold of these nanoresonators.

**Methods**

The optical cross-sections (Absorption Cross-Section: ACS, Scattering Cross-Section: SCS, Extinction Cross-Section: ECS) of the core-shell nanoparticle ($r_1$ = 25.5 nm inner and $r_2$ = 30 nm outer radii) and nanorod ($a_{long}$ = 75 nm long and $a_{short}$ = 25 nm short full axes) acting as a plasmonic nanoresonator at the central wavelength (795 nm) of the target illumination were determined by total-field-scattered-field method in Finite Element Method (FEM), namely via RF module of COMSOL Multiphysics (Fig. 1a, c, d). In case of the core-shell nanoparticle the thin shell composition was selected, since this results in an extinction dominated by the large absorption [4]. These nanoresonators were embedded into a polymer target, the index of refraction of this was determined by ellipsometry. The wavelength dependence of the real part of the refractive index was taken into account (at the central wavelength it is 1.54), whereas the imaginary part was neglected according to the negligible absorptance.

The analogous nanoresonator number densities ($\rho$) for core-shell nanoparticles and nanorods was determined based on the ratio of their ACS, by taking into account that the absorption coefficient is determined by the $\alpha$=ACS*$\rho$ relation, when low number density of nanoresonators is supposed.

To determine the absorptance and absorption coefficient, a supercell with typically 1 µm unit cell side-length was used, and the 21 µm long target was divided into at least 7 segments, each of 3 µm thick length along the pulse propagation direction, inside which the number density of nanoresonators was different (Fig. 1b). These supercells were illuminated by two counter-propagating linearly polarized short pulses injected through internal ports. At the side-boundaries periodic boundary condition was used, whereas behind the ports PML layers were inserted to eliminate spurious reflections. The transmittance and reflectance were extracted at interfaces inserted between the ports and PML layers, and during evaluation of the absorption coefficient the equivalence of the two opposite pulse propagation directions was supposed. The 21 µm target length corresponds to the full-width-at-half-maximum (FWHM) of a 120 fs pulse envelope. Such target is the most efficiently illuminated by the two counter-propagating pulses, when they overlap at the middle of the target at 240 fs.

The primary purpose is to improve the absorptance, accordingly the nanoresonator number density was varied in the $\rho$~1-15 µm$^{-3}$ interval. Primarily various distributions of the nanoresonators were tested, among them uniform (Fig. 2), single and double peaked Gaussian distributions (Fig. 3). Then *first* the nanoresonator distribution was optimized in steady-state RF module with the objective function of *minimal absorption difference in between neighbouring layers* (Fig 4, 5).

*Second* the steady-state absorptance and the average standard deviation (SD: δ) of absorptance sampled in different layers was determined by inspecting several random location distributions of the pre-optimized nanoresonator number density distribution. *Third* a typical system exhibiting a standard deviation of absorptance that approximates the average δ of the specific nanoresonator number density distribution was selected (Fig. 4-5/a). To diminish the difference in between symmetrically aligned layers embedding the same number of nanoresonators with different random location distributions, symmetrically arranged locations were supposed in analogous layers (A-G, B-F, C-E) (Fig. 1b). We can suppose that the location distribution possesses a mirror or even central symmetry without the loss of generality (Fig. 4, 5 inset).

*Fourth* the energy and energy density, as well as the power loss and power loss density distribution were analysed in different layers as a function of time (Fig. 4-5/b-e, (i) column). To ensure uniform e.g. phase transition or ignition the integrated energy deposited in different layers should be balanced. Accordingly, *fifth* the integrated energy and energy density as well as the power loss and power loss density distribution were determined throughout the maximal overlap of the counter-propagating pulses (Fig. 4-5/b-e, (ii) and (iii) column). *Finally*, the nanoresonator number density distribution was *successively re-adjusted* to ensure energy density standard deviation commensurate with (or even smaller than) the typical δ of absorption at 240 fs (Fig. 4-5/a-e, right column on histograms and right insets in subfigures).

To determine the time-evolution of the charge separation accompanying the short pulse illumination of the resonant core-shell nanoparticles and nanorods we implemented the hydrodynamic equations coupled with the Maxwell equations into the transient RF module of COMSOL Multiphysics (Fig. 6). The interaction of plasmonic nanoresonators with the 120 fs laser pulse was described via the wave equation including the induced polarization:

$$\nabla \times \nabla \times \vec{E}(\vec{r},t) + \frac{1}{c^2}\partial_{tt}\vec{E}(\vec{r},t) + \partial \mu_0 \partial_{tt}\vec{P}(\vec{r},t) = 0. \quad (1)$$

Here $\vec{E}(\vec{r},t)$ is the electric field with full spatio-temporal dependence, $\mu_0$ is the vacuum permeability and $\vec{P}(\vec{r},t)$ is the induced polarization. The time-dependence of the induced polarization was defined with the macroscopic position dependent electron density ($n_e$) and electron velocity ($v_e$) accounting for the polarization currents:

$$\partial_t \vec{P}(\vec{r},t) = -en_e\vec{v}_e \quad (2),$$

which couples the Maxwell equations to the hydro-dynamical equations as follows:

$$m_e\, n_e\, (\partial_t \vec{v}_e + \vec{v}_e \nabla \vec{v}_e) + \gamma\, m_e\, n_e\vec{v}_e = -en_e(\vec{E}\,\vec{v}_e \times \vec{H}), \quad (3)$$

$$\partial_t n_e + \nabla(n_e\vec{v}_e) = 0, \quad (4)$$

where *γ* is a phenomenological coefficient related to damping, while *e* and $m_e$ are the charge and mass of electron, respectively.

**Results**

*Absorptance cross-section of core-shell nanoparticles and nanorods resonant at the central wavelength of the illuminating pulse*

Via TFSF formulation the complete optical cross-section (OCS) of the nanoresonators was determined, including the absorption, scattering and extinction cross-section (ACS, SCS, ECS) (Fig, 1a, c, d). From the point of view of an uniform energy deposition inside target the absorption cross-section is the most relevant, accordingly this was compared for the core-shell and nanorod type nanoresonators.

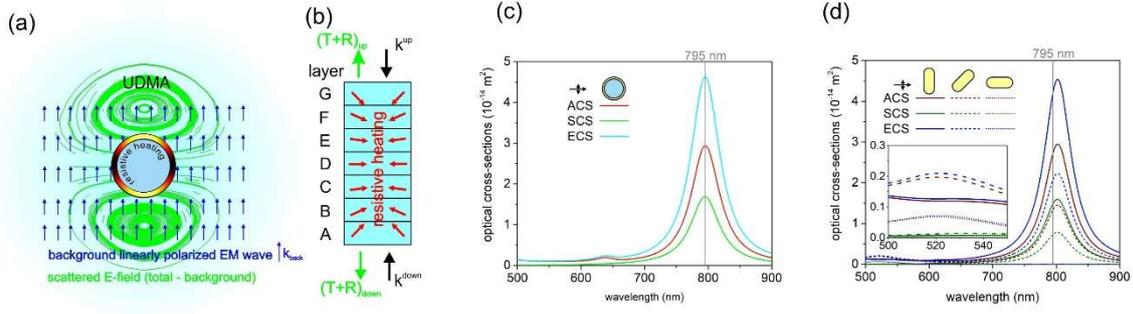

**Figure 1.** (a) TFSF formulation used to determine the OCS of nanoresonators. (b) Illumination of the targets by two counter-propagating pulses to extract the full optical response and to determine the absorptance. Orientation dependent OCS spectra of a resonant (c) core-shell nanoparticle and (d) nanorod in polymer medium.

The resonant core-shell nanoparticle exhibits $2.97*10^{-10}$ cm$^2$ cross-section, whereas the ACS of the resonant nanorod is $2.75/1.36/0.002/0.758*10^{-10}$ cm$^2$ in 0°/45°/90°/random orientation of the nanorod with respect to the **E**-field oscillation direction of the illuminating laser pulse. These values correspond to ACS$_{core-shell}$/ACS$_{nanorod}$ ratio of 1.08/2.18/1490/3.91, which indicates that the core-shell type nanoparticle exhibits a slightly larger ACS than the nanorod even in case of a longitudinal nanorod resonance. Accordingly, the core-shell nanoparticle is capable of resulting in ~4-times larger absorptance than the randomly oriented nanorods, when the same number density is applied inside the targets.

*Comparison of uniform nanoresonator distributions*

The illumination of a target containing uniform random distribution of nanoresonators resulted in an Ohmic loss, which decreases exponentially towards the middle of the sample, in accordance with the Lambert-Beer law. The Ohmic loss decreases symmetrically in case of illumination by two synchronously counter-propagating pulses, when the number density distribution of nanoresonators is symmetrical (Fig. 2).

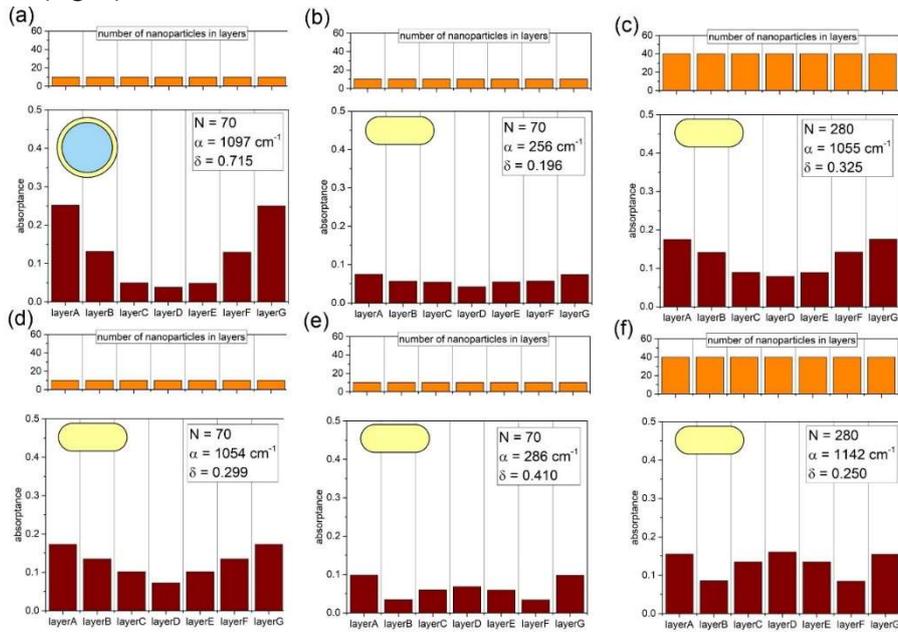

**Figure 2.** Uniform random distributions of (a) 70 core-shell nanoparticles and (d) 70 oriented nanorods, and (b,e-c,f) 70 - 280 randomly oriented nanorods.

The uniform random distribution of core-shell nanoparticles and oriented nanorods results in one single dip in the Ohmic loss at the centre of the target (Fig. 2a, d). In contrast, the uniform random distribution of randomly oriented nanorods promotes the appearance of one and two minima in the Ohmic loss as well (Fig. 2b, c and e, f). The appearance of one and double dips depends on the given location distribution. When two minima appear, they are symmetrically arranged with respect to the target centre.

The randomly oriented nanorods result in an absorptance, which is smaller than the absorptance achievable via core-shell nanoparticles, and the ratio of absorptances is in accordance with the ratio of the $ACS_{core-shell}/ACS_{nanorod}$, when the same number density is applied (Fig. 2a-to-b, d, e). Accordingly, by applying 4-times larger (280) number of nanorods the achieved absorptance becomes almost equal to that reached via (70) core-shell nanoparticles (Fig. 2a-to-c, f).

Based on the significantly larger and polarization independent absorptance the core-shell nanoresonators seem to be better choice, however the standard deviation of absorptance is several-times larger typically (Fig. 2a to b-f).

The illumination of nanorods aligned along the **E**-field oscillation direction resulted in an absorptance commensurate with that achievable via core-shell nanoparticles at a cost of standard deviation of absorptance, which becomes larger than the minimal δ achievable via randomly oriented nanorods (Fig. 2d-to-b, f). The random orientation of the nanorods promotes the minimization of the standard deviation of absorptance.

*Comparison of different Gaussian distributions*

Different Gaussian nanoresonator distributions were compared, which exhibit single or double peaks (Fig. 3). When the same number of nanoresonators was embedded, the ratio of the achieved absorptances was in accordance with the ratio of the $ACS_{core-shell}/ACS_{nanorod}$, independently of the distribution type. Accordingly, the 4-times larger (280) number of nanorods resulted in almost the same (1138 $cm^{-1}$ and 1092 $cm^{-1}$) absorption, as the (70) core-shell nanoparticles (1136 $cm^{-1}$ and 1102 $cm^{-1}$) (Fig. 3 a, b and c, d).

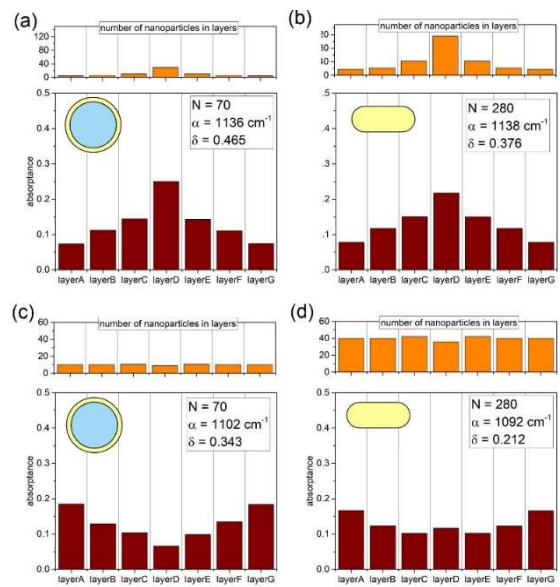

**Figure 3.** Comparison of different nanoresonator distributions: (a, b) wide single peaked Gaussian and (c, d) small amplitude double peaked Gaussian distribution of (a, c) core-shell and (b, d) nanorod nanoresonators.

However, the relation of the standard deviation of absorptance is nanoresonator distribution type dependent: it is smaller (i) for core-shell nanoparticles, when the single peaked Gaussian distribution is wider, (ii) for nanorods, when the double peaked Gaussian distribution has smaller amplitude.

In case of synchronized nanoresonator numbers resulting in commensurate absorptance both the wide single peaked and the small amplitude double peaked distribution exhibit a slightly larger standard deviation of absorptance for core-shells. The wide Gaussian distribution was selected as the primary distribution for steady-state optimization, since this makes it possible to reach larger absorption, although with a slightly larger $\delta$ of absorptance for both nanoresonators.

*Comparison of optimized and re-adjusted Gaussian distributions*

The energy and energy density as well as the power loss and power loss density were inspected in details for the nanoresonators in all layers as a function of time (Fig. 4, 5). When these quantities are analysed for both nanoresonators (for the core-shell nanoparticle), one has to consider the following: (i) Both the energy and power loss tendencies differ from the corresponding energy density and power loss density tendencies along the layers for either the instantaneous values, or for the values integrated throughout the specific time. This is due to that normalization is performed by division via the sum of nanoresonator volumes in layers, however this summarized volume differs in the neighbouring layers. (ii) The energy in the nanoparticle is smaller than in the full layer (but larger than in the shell), since the light energy is non-zero in the polymer (and in the silica core). (iii) The power loss in the nanoparticle equals with that in the full layer (moreover it equals with that in the shells), since the polymer (and the silica core) is non-absorbing. Accordingly, tendencies in energy and power loss as well as in their densities in nanoresonators are analogous for nanorods (differ for core-shells) along the layers, (however this difference is moderate due to the small volume fraction of the core).

The primarily optimized core-shell distribution determined by *minimizing the absorptance difference in between layers* consisted of 72 core-shell nanoparticles, and exhibited an $\alpha_{avg}$= 892 cm$^{-1}$ absorption coefficient, $A_{avg}$=0.846 average absorptance with $\delta_{avg}$=0.102 average standard deviation of absorptance (Fig. 4a: (i, ii) left columns). A typical random location distribution of the core-shell resonators exhibiting an $\alpha$= 958 cm$^{-1}$ absorption coefficient, A=0.866 absorptance with $\delta$=0.093 that are commensurate with the averaged values, was inspected further in time-domain (Fig. 4a: (iii)).

There is a well-defined difference in between energy (power loss) and energy density (power loss density) in different neighbouring layers, which increases with time (Fig. 4b (d) and c (e): (i, ii) left insets). At 240 fs delay corresponding to the complete overlap of both counter-propagating pulses at the target centre, the standard deviation of energy (power loss) and energy density (power loss density) is $\delta_{energy}$=0.387 ($\delta_{power\_loss}$=0.408) and $\delta_{energy\_density}$=0.541 ($\delta_{power\_loss\_density}$=0.567), respectively, which indicate 4.2 (4.4)- and 5.8 (6.1)-fold increase in uncertainty compared to that of the absorptance distribution (Fig. 4b (d) and c (e): (iii) left columns).

The successive modification of the number density distribution along the layers made it possible to improve the a core-shell nanoresonator doped target characteristics. For a typical system even larger $\alpha$= 1152 cm$^{-1}$ absorption coefficient, and larger A=0.911 absorptance with almost the same standard deviation of $\delta$=0.092 is reached, when compared to the wide Gaussian distribution determined via steady-state computations (Fig. 4a: right columns). The standard deviation of energy (energy density) is 8-times (almost 2-times) smaller, i.e. $\delta_{energy}$=0.048 ($\delta_{energy\_density}$=0.3) is achieved (Fig. 4b, c: (i, ii) right insets and (iii) right columns). The standard deviation of power loss (power loss density) is $\delta_{power\_loss}$=0.130 ($\delta_{power\_loss\_density}$=0.371), which is approximately 3-times (1.5-times) smaller than the values in case of the primarily optimized distribution (Fig. 4d, e:(i, ii) right insets and (iii) right columns).

The integrated quantities rapidly increase in time in the adjusted core-shell distribution as well (Fig. 4 b-e: (ii, iii) left-to-right insets and columns). However, due to the successively minimized integrated energy difference, the rate of integrated energy increase is similar in neighboring layers, which ensure more uniform energy dissipation throughout the complete target (Fig. 4b: (ii, iii), right inset and columns). For the energy density the difference in time-dependence and the resulted $\delta$ is significantly larger according to the different core-shell nanoparticle number density in the neighboring layers (Fig. 4b-to-c: (i, ii), right insets). The largest energy density difference appears between the central layers of the target, where the nanoparticle density is the highest (Fig. 4c: (iii), right columns). The power loss and power loss density exhibit similar tendencies according to the slight difference related to the energy in the core (Fig. 4 d, e).

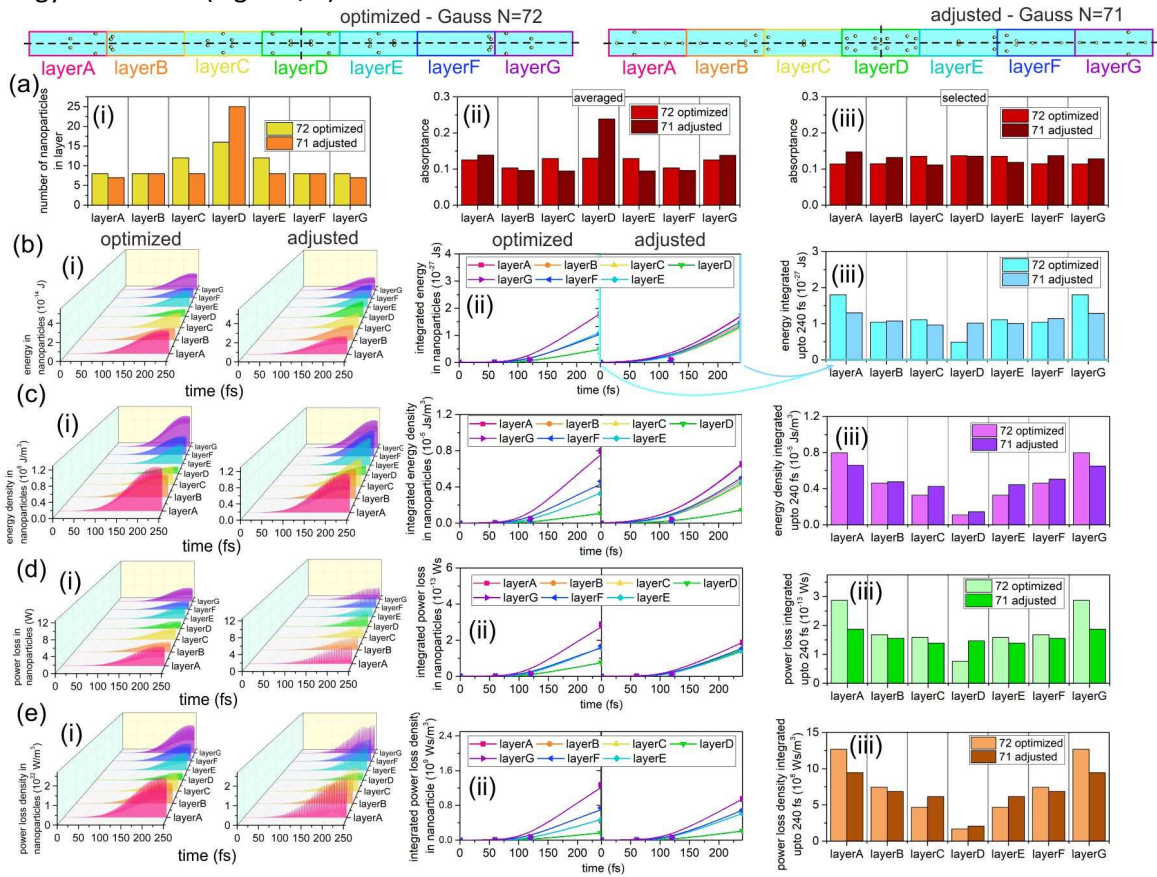

**Figure 4.** (a) (i) Centrally symmetric primarily optimized and adjusted core-shell Gaussian distribution, and the corresponding (ii) absorptance distribution of several random systems, and (iii) of a typical system with a standard deviation approximating the average $\delta$ of absorptance. (b / c / d / e) The corresponding (i) core-shell nanoparticle energy / energy density / power loss / power loss density distribution in different layers as a function of time and (ii) integrated throughout different time-intervals and (iii) the distribution of values integrated in different layers at 240 fs. Inset: schematics of the symmetric core-shell distributions in the target.

The primary nanorod distribution was determined similarly by *minimizing the absorptance difference in between layers* (Fig. 5a: (i, ii), left columns). The primarily optimized target consisted of 288 nanorods and exhibited $\alpha_{avg}$= 948 cm$^{-1}$ absorption coefficient, A$_{avg}$=0.863 average absorptance with $\delta_{avg}$=0.137 average standard deviation. First a typical random location distribution exhibiting $\alpha$= 1175 cm$^{-1}$ absorption coefficient, A=0.915 absorptance with $\delta$=0.162 that are slightly larger than the averaged values, was inspected in time-domain (Fig. 5a: (iii), left columns). The absorption coefficient and absorptance is in accordance with the expectation: the four-times larger number of nanorods and smaller ACS result in almost the same values, as in the core-shell distribution.

There is an obvious difference in between energy (power loss) and energy (power loss) density in different neighbouring layers, which increases with time, similarly to the core-shell distribution (Fig. 5b (d) and c (e): (i, ii) left insets).

At 240 fs delay corresponding to the complete overlap of the two counter-propagating pulses at the target centre, the standard deviation of energy and power loss is $\delta_{energy}=\delta_{power\_loss}=0.633$, whereas the standard deviation of energy and power loss density is $\delta_{energy\_density}=\delta_{power\_loss\_density}=0.535$, which indicate 3.9- and 3.3-fold increase in uncertainty compared to that of the absorptance distribution respectively (Fig. 5b (d) and c (e): (iii) left columns). There is no difference neither in between tendencies of energy and powerloss nor in between energy density and power loss density, since the complete energy is transformed to power loss in the compact metal nanorod.

The proposed nanorod distribution was determined re-adjusting the previously optimized Gaussian distribution. A typical system achieved via the successive modification of the number density distribution along the layers made it possible to reach a similar $\alpha= 987$ cm$^{-1}$ absorption coefficient and A=0.874 absorptance (Fig. 5a: right columns). However, this can be achieved as a cost of almost four-times larger $\delta$=0.605 standard deviation of absorptance. The advantage of the successively adjusted distribution is that the standard deviation of energy and power loss (energy density and power loss density) is 7.7-times (2.5-times) smaller compared to the values in case of the primarily optimized distribution, namely, $\delta_{energy}$=0.082 ($\delta_{energy\_density}$=0.212) is achieved (Fig. 5b, c and d, e: (i, ii) right insets and (iii) right columns).

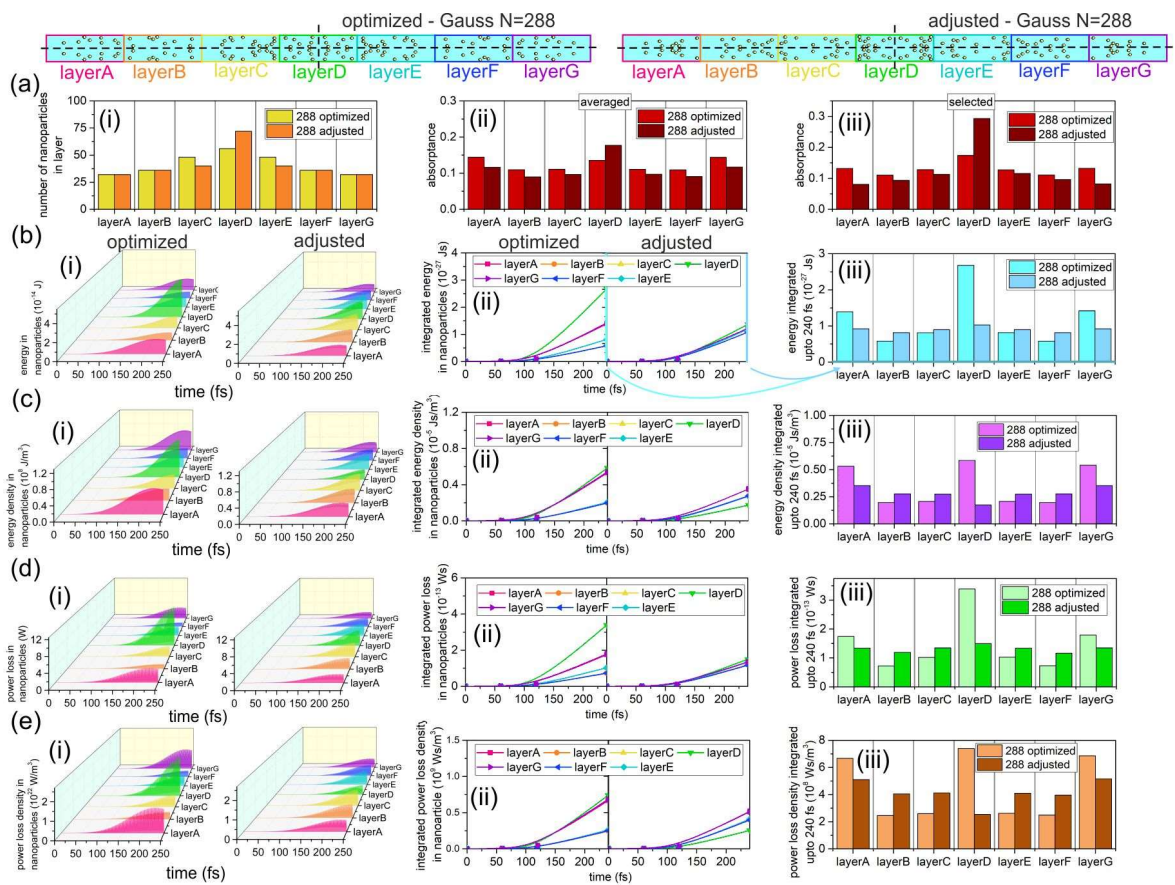

**Figure 5.** (a) (i) Centrally symmetric primarily optimized and adjusted nanorod Gaussian distributions, and the corresponding (ii) absorptance distributions of several random systems, and (iii) of a typical system with a standard deviation approximating the average $\delta$ of absorptance. (b / c / d / e) The corresponding (i) nanorod energy / energy density / power loss / power loss density distribution in different layers as a function of time and (ii) integrated throughout different time-intervals and (iii) the distribution of values integrated in different layers at 240 fs. Inset: schematics of the symmetric nanorod distributions in the target.

The tendencies of integrated energy, power loss, and their corresponding densities in the successively approximated nanorod distribution is moderated with respect to the primarily optimized one similarly to the counterpart core-shell distributions presented earlier (Fig. 4 b-e: left-to-right insets and columns). The only characteristic difference between the core-shell and nanorod distributions is that the power loss and power loss density exhibit the same tendencies as energy and energy density according to the proportionality of the energy and power loss in the nanorod, therefore their standard deviation is analogous (Fig. 5b-to-d, c-to-e). Moreover, not only the trends but the instantaneous and integrated values are also comparable for all of the distributions examined, despite the four-fold difference between the densities of core-shell nanoparticles and nanorods. This again indicates the governing role of the nanoresonator ACS at a given central wavelength. It seems to be counterintuitive that the corresponding density values are also close to each other, but this is the result of the five-times larger total core-shell volume. However, the difference in standard deviation stems only partially from the differences in the number density distributions along the layers, it is affected by the actual nanoresonator positions inside layers as well.

A direct comparison of the energy, power loss and their density distributions of the core-shell and nanorod doped targets uncovers that while the core-shell distributions show a minimum in all these quantities at the target center, only the adjusted nanorod distributions exhibit minima but only in the energy and power loss densities as a result of normalization by the large summarized nanoparticle volume at the central layer (Fig. 4 b-e (iii) to Fig. 5 b-e (iii)). According to the central symmetry of the inspected core-shell nanoparticle and nanorod systems there is a slight difference in either of the absorptance, energy (power loss), moreover in the energy density (power loss density) distribution in corresponding symmetrically aligned layers (Fig. 4 and 5).

In case of the proposed nanorod distribution the absorption coefficient and the absorptance is just slightly smaller, the standard deviation of absorptance is 6.6-times larger, whereas $\delta$ is 1.7-times larger (1.4-times smaller) than the energy (energy density) standard deviation in case of the proposed core-shell distribution. Further advantages are that standard deviation of power loss (power loss density) is 1.5 (1.8) times smaller in case of the optimized nanorod distribution.

*Charge separation study just below the damage threshold*

Our results prove that the time-dependent charge separation computed via a hydrodynamic model is very similar to the charge separation determined by a linear model, which confirms that the absorption and absorptance is almost linearly scaled with the intensity throughout the damage threshold.

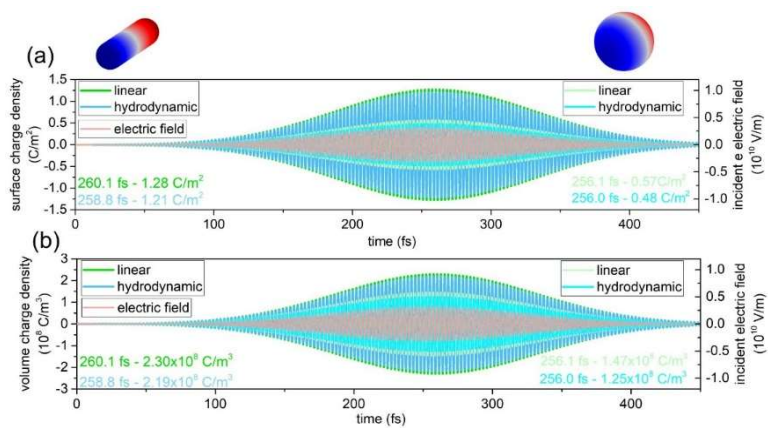

**Figure 6.** (a) Surface and (b) corresponding volume charge separation accompanying the resonance induced on the core-shell nanoparticle and nanorod illuminated by 120 fs short pulse of 795 nm central wavelength at the damage threshold.

The 25 nm*75 nm resonant nanorod supports a charge oscillation, which reaches a maximum of 1.28 C/m$^2$ (1.21 C/m$^2$) in surface charge separation corresponding to a maximum of 2.30*10$^8$ C/m$^3$ (2.19*10$^8$ C/m$^3$) in volume charge separation at 260.1 fs (258.8 fs) based on the linear (hydrodynamic model). The 25.5 nm / 30 nm resonant core-shell nanoparticle exhibits a charge oscillation of smaller amplitude, which reaches a maximum of 0.57 C/m$^2$ (0.48 C/m$^2$) in surface charge separation corresponding to 1.47*10$^8$ C/m$^3$ (1.25*10$^8$ C/m$^3$) in volume charge separation at 256.1 fs (256.0 fs) based on the linear (hydrodynamic model).

**Conclusion**

Considering that the absorptance is orientation independent and significantly larger in case of the same number of nanoresonators, the core-shell nanoparticles turned to be a better choice based on steady-state computations. Although, the standard deviation of absorptance was larger in the uniform and in different predefined Gaussian distributions for core-shell nanoresonators, it became smaller in the primarily optimized distributions (except for energy and power loss densities). In case of the adjusted distributions the standard deviation of the absorptance and energy was smaller for core-shell nanoresonators, but became larger for energy density as well as for power loss and power loss density. The users' purposes define, whether core-shell nanoparticles possessing larger ACS and resulting in larger absorptance are preferred or nanorods, which distributions exhibit a compromised standard deviation but are more easily synthesizable via simple chemical procedures.

**Acknowledgement**

This work was supported by the National Research, Development and Innovation Office (NKFIH) of Hungary, projects: "Optimized nanoplasmonics" (K116362), "Ultrafast physical processes in atoms, molecules, nanostructures and biological systems" (EFOP-3.6.2-16-2017-00005) and the Eötvös Lóránd Research Network of Hungary. The authors acknowledge the helpful discussions with Professor Pavel Ginzburg on hydrodynamic modelling of plasmonic nanostructures. Balázs Bánhelyi acknowledges support by the János Bolyai Research Scholarship of the Hungarian Academy of Sciences. Emese Tóth acknowledges support by the project "Integrated program for training new generation of scientists in the fields of computer science", (EFOP-3.6.3-VEKOP-16-2017-00002).

**References**

[1] E. Prodan and P. Nordlander, Plasmon hybridization in spherical nanoparticles, J. Chem. Phys. 120, 5444 (2004)

[2] Halas, N. Playing with Plasmons: Tuning the Optical Resonant Properties of Metallic Nanoshells. MRS Bulletin 30, 362–367 (2005)

[3] R. Bardhan, N. K. Grady, T. Ali, and N. J. Halas, Metallic Nanoshells with Semiconductor Cores: Optical Characteristics Modified by Core Medium Properties, ACS Nano, 4, 10, 6169–6179 (2010).

[4] F. Tam, A. L. Chen, N. J. Halas, Mesoscopic nanoshells: Geometry-dependent plasmon resonances beyond the quasistatic limit, J. Chem. Phys. 127, 204703 (2007)

[5] S.J. Oldenburg, R.D. Averitt, S.L. Westcott, N.J. Halas, Nanoengineering of optical resonances, Chemical Physics Letters 288 243–247 (1998)

[6] W. Ni, X. Kou, Z. Yang, and J. Wang, Tailoring Longitudinal Surface Plasmon Wavelengths, Scattering and Absorption Cross Sections of Gold Nanorods ACS Nano, 2, 4, 677–686 (2008)

[7] V. Juvé et. al., Size-Dependent Surface Plasmon Resonance Broadening in Nonspherical Nanoparticles: Single Gold Nanorods, Nano Lett. 13, 2234–2240 (2013)

[8] S. Lal, S. Link, N. Halas, Nano-optics from sensing to waveguiding. Nature Photon 1, 641–648 (2007)

[9] P. Zijlstra, P. Paulo, M. Orrit, Optical detection of single non-absorbing molecules using the surface plasmon resonance of a gold nanorod. Nature Nanotech 7, 379–382 (2012).

[10] A. Szenes, B. Bánhelyi, L. Zs. Szabó, G. Szabó, T. Csendes, M. Csete, Improved emission of SiV diamond color centers embedded into concave plasmonic core-shell nanoresonators, Sci Rep 7, 13845 (2017)

[11] A. Szenes, B. Bánhelyi, L. Zs. Szabó, G. Szabó, T. Csendes, M. Csete, Enhancing Diamond Color Center Fluorescence via Optimized Plasmonic Nanorod Configuration, Plasmonics 12, 1263–1280 (2017)

[12] A. M. Gobin et. al., Near-Infrared Resonant Nanoshells for Combined Optical Imaging and Photothermal Cancer Therapy, Nano Lett. 7, 7, 1929–1934 (2007)

[14] W. I. Choi et al., Tumor Regression In Vivo by Photothermal Therapy Based on Gold-Nanorod-Loaded, Functional Nanocarriers, ACS Nano, 5, 3, 1995–2003 (2011)

[15] Kern, C. et al. Comparison of femtosecond laser-induced damage on unstructured vs. nano-structured Au-targets. *Appl. Phys. A* **104**, 15–21 (2011).

[16] B. J. Nagy et al.: "Near-Field-Induced Femtosecond Breakdown of Plasmonic Nanoparticles" *Plasmonics* **15**, 335–340 (2020)

[17] A. Plech et. al.: Femtosecond laser near-field ablation from gold nanoparticles *Nature Physics* **2**, 44–47 (2006)

[18] B. Rethfeld et. al: Ultrafast dynamics of nonequilibrium electrons in metals under femtosecond laser irradiation *Physical Review B*, **65**, 214303 (2002)

[19] A.V. Krasavin et al. Nonlocality-driven supercontinuum white light generation in plasmonic nanostructures *Nature Communications* **7**, 11497 (2016)

[20] P. Ginzburg, Nonlinearly coupled localized plasmon resonances: Resonant second-harmonic generation, *Physical Review B* **86**, 085422 (2012)

[21] P. Ginzburg et al.: Nonperturbative Hydrodynamic Model for Multiple Harmonics Generation in Metallic Nanostructures, ACS Photonics 2 (2015)

[22] M. Scalora et al.: "Second and third-harmonic generation in metal-based structures" Physical Review A 82 (2010)

[23] K. Tanabe: "Plasmonic energy nanofocusing for high-efficiency laser fusion ignition", JJAP 55 (2016) 08RG01

[24] K. Tanabe: "Plasmonic Concepts for Condensed Matter Nuclear Fusion", JCMNS 24 (2017) 296-300.

[25] Purvis et al.: "Relativistic plasma nanophotonics for ultrahigh energy density physics", Nature Photonics (2017) 796-800.

[26] V. Kaymak et al.: "Nanoscale Ultradense Z-Pinch Formation from Laser-Irradiated Nanowire Arrays" PRL 117 (2016) 035004

[27] C. Bargsten et al.: "Energy penetration into arrays of aligned nanowires irradiated with relativistic intensities: Scaling to terabar pressures", Science Advances 3 (2017) 1601558.

[28] L. P. Cserani et al.: "Radiation dominated implosion with nano-plasmonics", Laser and Particle Beams (2018) 1–8.

[29] L.P. Csernai et al.: "Radiation-Dominated Implosion with Flat Target", Phys. Wave Phen. 28, (2020) 187–199.

[30] I. Papp et al.: "Laser wake field collider", Phys. Lett. A 396 (2021) 127245.